\documentclass[twocolumn,showpacs,preprintnumbers,amsmath,amssymb]{revtex4}
\usepackage{graphicx}
\usepackage{dcolumn}
\usepackage{bm}

\begin{document}


\def\KONYA{Department of Physics, University of Sel\c{c}uk, 42079 Konya,
Turkey}
\def\FIAS{Frankfurt Institute for Advanced Studies, J.W. Goethe University,
D-60438 Frankfurt am Main, Germany}
\def\MOSCOW{Institute for Nuclear Research, Russian Academy of Sciences, 117312 
Moscow, Russia}

\newcommand{\goo}{\,\raisebox{-.5ex}{$\stackrel{>}{\scriptstyle\sim}$}\,}
\newcommand{\loo}{\,\raisebox{-.5ex}{$\stackrel{<}{\scriptstyle\sim}$}\,}

\title{INFLUENCE OF ANGULAR MOMENTUM AND COULOMB INTERACTION OF COLLIDING 
NUCLEI ON THEIR MULTIFRAGMENTATION}

\affiliation{\KONYA}
\affiliation{\FIAS}
\affiliation{\MOSCOW}

\author{A.~Ergun}    \affiliation{\KONYA}
\author{H.~Imal}    \affiliation{\KONYA}
\author{N.~Buyukcizmeci}    \affiliation{\KONYA}
\author{R.~Ogul}   \affiliation{\KONYA}
\author{A.S.~Botvina}      \affiliation{\MOSCOW}\affiliation{\FIAS}

\date{\today}

\begin{abstract}
Theoretical calculations are performed to investigate the angular momentum 
and Coulomb effects on fragmentation and multifragmentation in peripheral 
heavy-ion collisions at Fermi energies. Inhomogeneous distributions of hot 
fragments in the 
freeze-out volume are taken into account by microcanonical Markov chain
calculations within the Statistical Multifragmentation Model (SMM). 
Including an angular momentum and a long-range Coulomb interaction between 
projectile and target residues leads to new features in the statistical 
fragmentation picture. In this case, one can obtain specific correlations 
of sizes of emitted fragments with their velocities and an emission in the 
reaction plane. In addition, one may see a significant 
influence of these effects on the isotope production both in the midrapidity 
and in the kinematic regions of the projectile/target. The relation of this 
approach to the simulations of such collisions with dynamical models is also 
discussed. 
\end{abstract}


\maketitle
\section{Introduction}

It has been commonly accepted since long ago that in central heavy-ion 
collisions at Fermi energies (20--50 MeV per nucleon) relatively high 
excitation energies of nuclear matter, with temperatures up to 
$T\approx$5--8 MeV, can be reached \cite{dagostino96}. 
Therefore, they become a suitable 
tool to study the equation of state (EoS) of hot nuclear matter 
and the nuclear liquid-gas phase transitions at subnuclear densities. 
As discussed previously 
\cite{Botvina02}, with the help of multifragmentation 
one can study the properties 
of hot fragments in the vicinity of other nuclear species. 
The angular momentum effect is usually disregarded in this case, 
since the impact parameters are small. 
During the peripheral heavy-ion collisions at the same energies, 
a considerable amount of angular momentum could be transferred from 
the interaction region to the excited projectile and target residual 
nuclei, and this can lead to significant changes in their 
multifragmentation \cite{BotvinaGross95,Gross97,Botvina01}. 
Additional long-range forces caused by the complicated Coulomb 
interaction between the target and projectile-like 
sources are involved essentially in the process \cite{Botvina99,Botvina01}: 
The multifragmentation in the presence of the external Coulomb field offers 
a possibility to study, experimentally, the effects of this 
long-range force, which are very important for disintegration of matter 
\cite{Gross97}. This is also necessary for construction of a reliable EoS 
of nuclear matter at subnuclear 
densities. Another motivation of these studies is that similar 
conditions for nuclear matter take place during the collapse and explosion of 
massive stars and in the crust of neutron stars~\cite{Lattimer01,Botvina10}, 
where the Coulomb interactions of dense electron environment change the 
fragmentation picture. 
It is generally assumed that the statistical equilibrium regarding the 
fragment composition at subnuclear densities should be established in 
these astrophysical cases. Therefore, the analysis of the observables obtained 
in laboratory experiments with 
statistical models is a proper way to get knowledge on stellar matter. 
Previous studies of isospin composition of the produced fragments were 
found to be especially important for determining the strength of the 
symmetry energy during fragment formation in hot and diluted 
environments~\cite{Botvina02,Ono03,LeFevre05,Ogul11}. 
 
In the analysis of ALADIN data, charge and isotope 
yields, fragment multiplicities and temperatures, and correlations of various 
fragment properties were successfully described by statistical ensemble 
approach ~\cite{Ogul11,Buyukcizmeci05,Botvina06,botvina92,botvina95,xi97} 
within the Statistical 
Multifragmentation Model (SMM) ~\cite{bondorf95}. This was also achieved in 
the analysis of the experimental data of Liu et al.~\cite{Liu04} obtained 
at the MSU laboratory at 50 MeV/nucleon 
\cite{Ogul09,Buyukcizmeci11,Buyukcizmeci12}, and in the analysis of TAMU 
data ~\cite{Iglio06,Souliotis07}. In these studies, the symmetry energy 
of fragments was one of the main model parameters 
governing the mean $N/Z$ values, the isoscaling parameters, and the isotopic 
composition of the fragments. 
For interpretation of ALADIN and MSU experiments, which can be explained 
by formation and decay of single thermalized sources, we considered the 
averaged Coulomb interaction of fragments (the Wigner-Seitz approximation), 
since a direct positioning of fragments in the freeze-out volume 
has minor influence on their charge and isotope distributions. 
This is well justified for relativistic peripheral collisions, 
and central collisions of heavy nuclei around the Fermi energy. 
However, important information on multifragmentation and properties of 
fragments can be extracted in peripheral collisions at Fermi energies as well. 
The new fragment partitions can be obtained by including the Coulomb effects 
caused by the proximity of colliding target and projectile nuclei, as well 
as the effects by large angular momentum 
transfer to the multifragmentating sources. For example, 
a long-range Coulomb interaction of the target- and projectile-like 
sources changes the fragmentation pattern and leads to a predominant 
midrapidity (neck--like) emission of light 
and intermediate mass fragments (IMF, with charge numbers $Z=$3--20). Few such 
experiments have already been analyzed with 
statistical models \cite{Jandel05,Souliotis07}. 
However, there were no systematic theoretical investigations of the Coulomb 
and angular momentum effects on multifragmentation picture at these 
reactions, especially on the isotope yields which are crucial for 
astrophysical applications. As it was suggested in 
Ref.~\cite{Botvina01,Botvina99}, the angular momentum may lead to 
more neutron-rich IMF production and to anisotropic emission with respect to 
the projectile and target sources. 

In this paper, we shall theoretically investigate the influence of angular 
momentum and Coulomb interactions on the charge yields, the neutron to proton 
ratios and the velocity distributions of hot particles for 
peripheral $^{84}$Kr+$^{84}$Kr collisions at 35 MeV per nucleon. 
We believe that this is a quite typical reaction, and our selection is partly 
motivated by recent FAZIA experiments \cite{fazia13}. 
For the simulation of the reactions, we consider the break-up of a single 
source $^{84}$Kr in the proximity of a secondary source $^{84}$Kr, as a 
symmetric system in terms of isospin contents. The calculations are 
carried out within the Markov chain version of the statistical 
multifragmentation model, which is designed for a microcanonical simulation 
of the decay modes of nuclear sources \cite{Botvina01,Botvina00}. 
This method is based on producing the Markov chain of partitions which 
characterize the whole statistical ensemble. In this method the individual 
fragment partitions and coordinate positions of fragments in the 
freeze-out volume are generated. They are selected by the Metropolis 
algorithm and we can take into account the influences of angular momentum 
and Coulomb interactions for each spatial configuration of primary 
fragments in the freeze-out volume, similar to 
Refs.~\cite{BotvinaGross95,Gross97}.

\begin{figure} [tbh]
\begin{center}
\includegraphics[width=8.6cm,height=11cm]{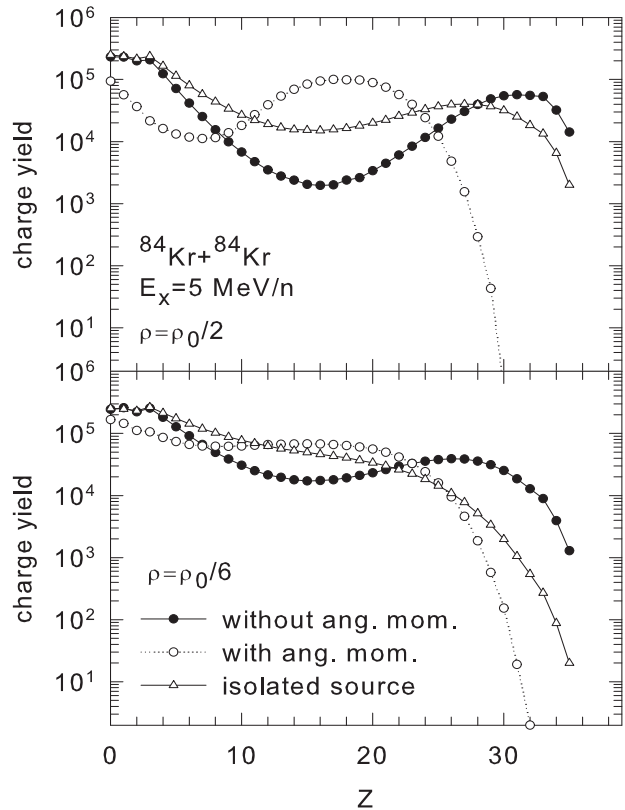}
\end{center}
\caption{\small{Total charge yield of primary hot fragments, in the cases of 
without (full circles) and with angular momentum  (open circles, $L=80\hbar$), 
after multifragmentation of the projectile $^{84}$Kr source at the excitation 
energy $E_x=5$ MeV/nucleon. This source is assumed to be formed 
in the peripheral $^{84}$Kr + $^{84}$Kr collision at 35 MeV/nucleon, and its 
disintegration is affected by the Coulomb field of the target source. 
Top and bottom panels show the results at freeze-out densities 
$\rho=\rho_0/2$, and $\rho=\rho_0/6$, respectively. For comparison, the 
results of multifragmentation of a single isolated $^{84}$Kr source, at the 
same excitation energy but without the external Coulomb field and without 
angular momentum, are shown too.}}
\end{figure}
\begin{figure} [tbh]
\begin{center}
\includegraphics[width=8.6cm,height=8.6cm]{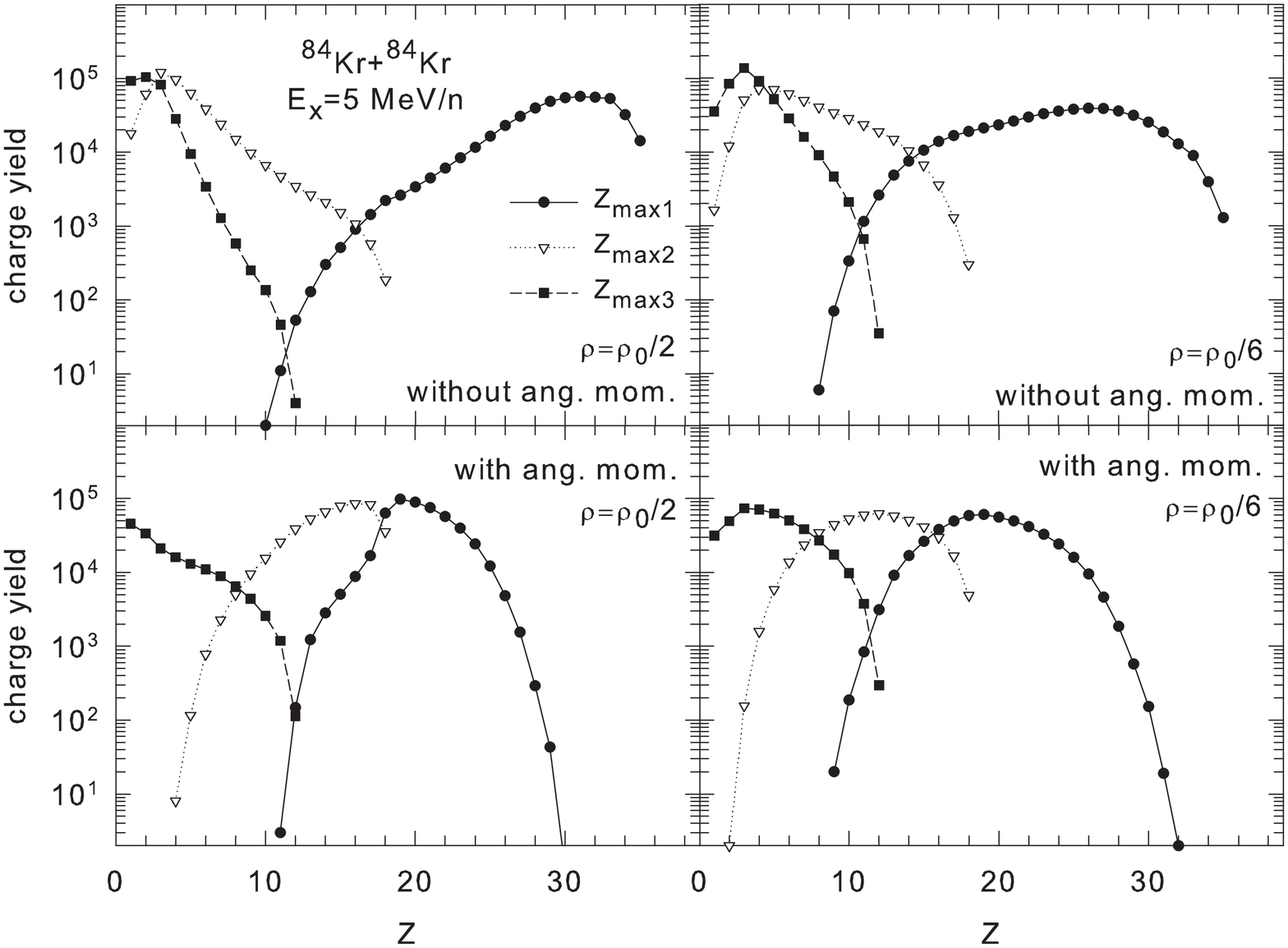}
\end{center}
\caption{\small{ Charge yield of the first, second and third largest hot 
fragments ($Z_{max1},Z_{max2}$, and $Z_{max3}$) after multifragmentation 
of the $^{84}$Kr source (as in Fig.~1). Top panels show the results without 
angular momentum, and bottom panels with angular momentum ($L=80\hbar$).}}
\end{figure}
\section{Statistical Approach To Multifragmentation}

It is assumed in the microcanonical SMM 
that a statistical equilibrium is reached at low density 
freeze-out region. The breakup channels are composed of nucleons and 
nuclear fragments, and the laws of conservation of energy $E_x$, momentum, 
angular momentum, mass number $A$ and charge number $Z$ are considered. 
Besides the breakup channels, the compound-nucleus channels are also 
included, and competition between all channels is permitted. In this way, 
the SMM covers the conventional evaporation and fission processes 
occurring at low excitation energy as well as the transition region 
between the low and high energy de-excitation regimes. In the thermodynamic 
limit, SMM is consistent with liquid-gas phase transitions when the 
liquid phase is represented by infinite nuclear clusters \cite{Das98}, 
that allow connections for the astrophysical cases \cite{nihal13}. 
We calculate the statistical  weights of all breakup channels partitioning 
the system into various species. The decay channels are generated by 
Monte Carlo method according to their statistical weights. 
In the Markov chain SMM \cite{Botvina01,Botvina00} 
we use also ingredients taken from the standard 
SMM version developed in the Refs.~\cite{botvina85,botvina87,bondorf95} 
which was successfully used for comparison with various experimental data: 
Light fragments with mass number $A\leq 4$ and charge number $Z\leq 2$ 
are considered as 
elementary particles with the corresponding spins (nuclear gas) that have 
translational degrees of freedom. The fragments with mass number $A>4$ are 
treated as heated nuclear liquid drops. In this way one can study the 
nuclear liquid-gas coexistence in the freeze-out volume. Free energies 
$F_{A,Z}$ of each fragment are parameterized as a sum of the bulk, 
surface, Coulomb and symmetry energy contributions 
\begin{equation}
F_{A,Z}=F_{A,Z}^B+F_{A,Z}^S+E_{A,Z}^C+E_{A,Z}^{sym}.
\end{equation}
The bulk contribution is given by $F_{A,Z}^B=(-W_0-T^2/
\varepsilon_0)A$, where $T$ is the temperature, the parameter $
\varepsilon_0$ is related to the level density, and $W_0=16$ MeV
is the binding energy of infinite nuclear matter. 
Contribution of
the surface energy is $F_{A,Z}^S=B_0 A^{2/3}[(T_{\rm
c}^2-T^2)/(T_{\rm c}^2+T^2)]^{5/4}$, where $B_0=18$ MeV is the
surface energy term, and $T_{\rm c}=18$ MeV the critical
temperature of the infinite nuclear matter. 
In the standard SMM version the Coulomb energy 
contribution is $E_{A,Z}^C=cZ^2/A^{1/3}$, where c denotes the
Coulomb parameter obtained in the Wigner-Seitz approximation,
$c=(3/5)(e^2/r_0)(1-( \rho / \rho_0)^{1/3})$, with the charge unit
e, $r_0=1.17$ fm, and $\rho_0$ is the normal nuclear matter
density (0.15 fm$^{-3}$). 
However, within this Markov-chain SMM we directly calculate the 
Coulomb interaction of non-overlapping fragments in the freeze-out 
by taking into account their real coordinate positions. 
The symmetry term is
$E_{A,Z}^{sym}= \gamma (A-2Z)^2/A$, where $\gamma =25$ MeV is the
symmetry energy parameter. All the parameters given above are
taken from the Bethe-Weizsaecker formula and correspond to the
assumption of isolated fragments with normal density 
unless their modifications in the hot and dense freeze-out configuration 
follow the analysis of experimental data. For the freeze-out density, 
one-third of normal nuclear matter density is assumed in many successful 
studies and consistent with independent experimental determination in 
sources formed in peripheral nuclear collisions \cite{Fritz99,Viola04}. 
To be more general, in this work we use $\rho=\rho_0/2$ and 
$\rho=\rho_0/6$ densities for better evaluation of Coulomb and angular 
momentum effects. The different positioning of particles and volume 
parameters is also useful for understanding the origin of the kinetic 
energies of fragments observed in experimental data. 
In the case of the large density ($\rho=\rho_0/2$) we assume a deformation 
of fragments in the freeze-out volume by effectively reducing distances 
between the fragments  for calculation of their Coulomb interaction 
by a factor of 0.7. This can be partly justified by non-spherical shape 
of these fragments since they are excited. Usually, we generate about $5.10^5$ 
Monte Carlo events to provide sufficient statistics. 
\begin{figure} [tbh]
\begin{center}
\includegraphics[width=8.6cm,height=11cm]{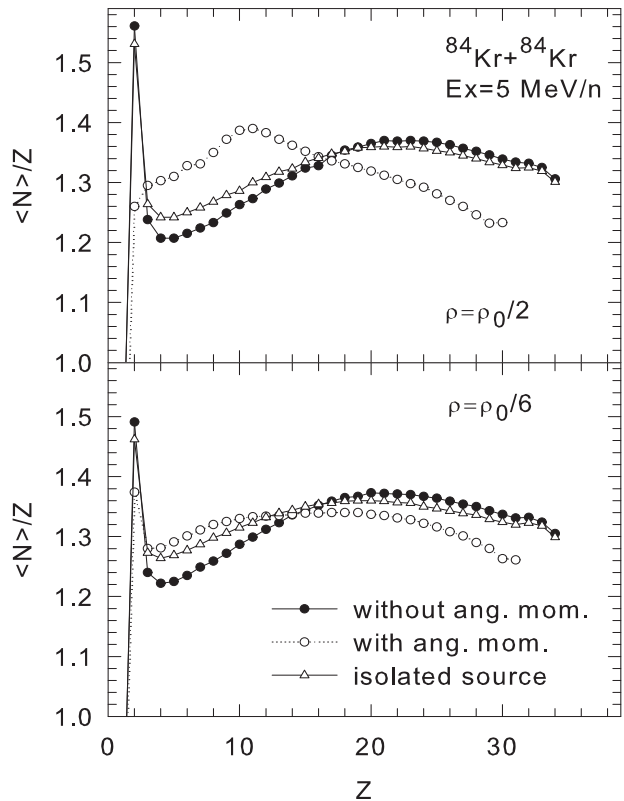}
\end{center}
\caption{\small{The neutron-to-proton ratio $<N>/Z$ of hot primary fragments 
produced at the freeze-out density $\rho=\rho_0/2$ (top panel) and 
$\rho=\rho_0/6$ (bottom panel). Other notations are as in Fig.~1.}}

\end{figure}
\begin{figure} [tbh]
\begin{center}
\includegraphics[width=8.6cm,height=11cm]{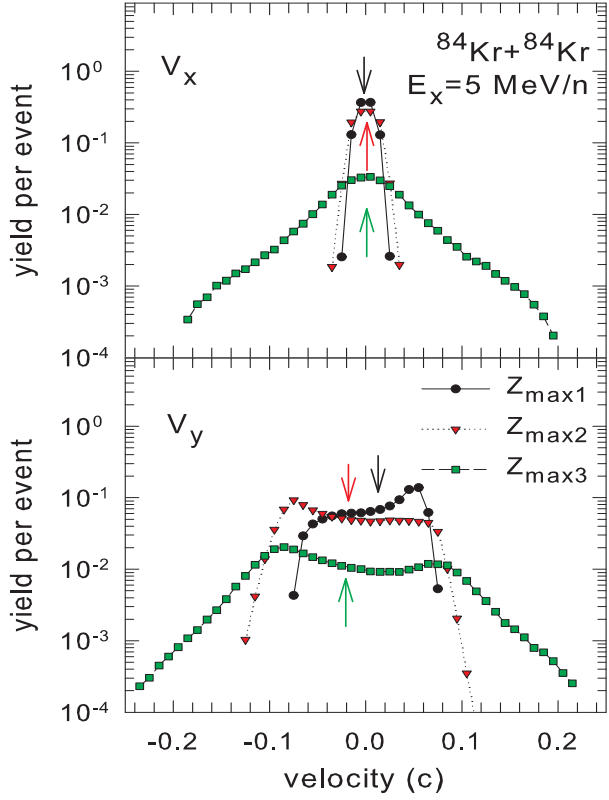}
\end{center}
\caption{\small{(Color online) $V_x$ and $V_y$ velocity distributions of 
the first, second and third largest fragments in multifragmentation of 
$^{84}$Kr with the angular momentum $L=80\hbar$ and at the freeze-out 
density $\rho=\rho_0/2$.}}
\end{figure}
\begin{figure} [tbh]
\begin{center}
\includegraphics[width=8.6cm,height=11cm]{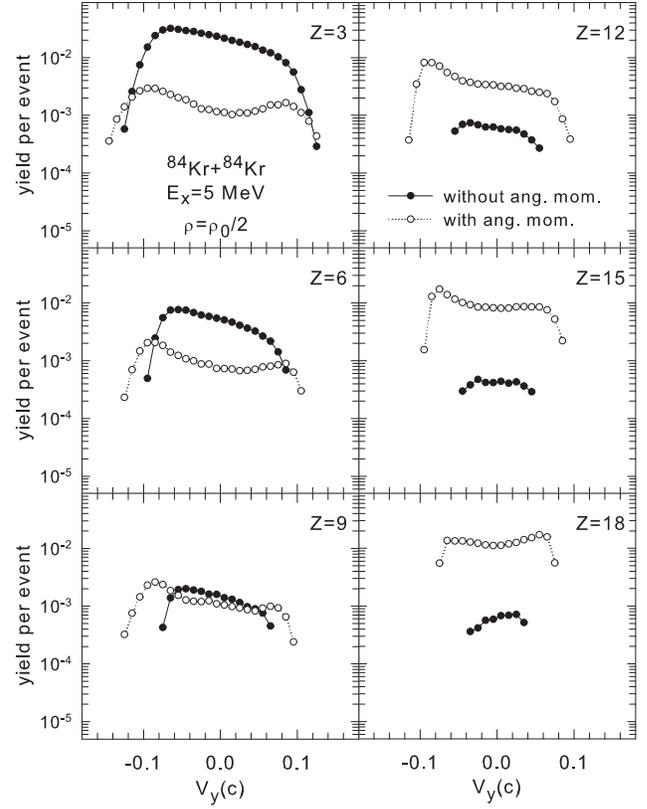}
\end{center}
\caption{\small{The relative yields of specific fragments (see panels) 
coming from the disintegration of $^{84}$Kr projectile source at the 
excitation energy of 5 MeV/nucleon as a function of the velocity $V_y$ 
in the source frame. The calculations have been performed with (open circles 
for $L=80\hbar$), and without angular momentum (full circles) at the 
freeze-out density $\rho=\rho_0/2$.}}

\end{figure}
\begin{figure} [tbh]
\begin{center}
\includegraphics[width=8.6cm,height=11cm]{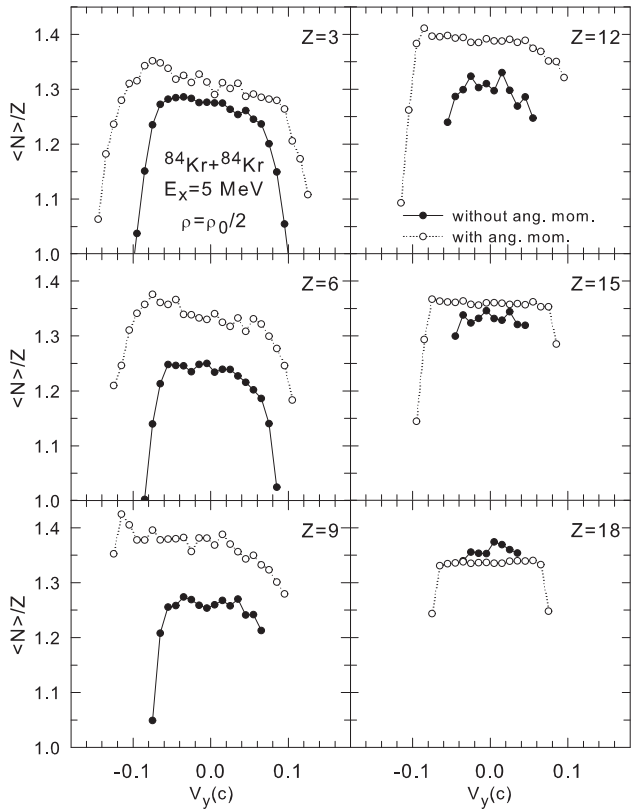}
\end{center}
\caption{\small{Average neutron to proton ratios as a function of $V_y$
shown in Fig.~5 (notations are the same).}}
\end{figure}
\begin{figure} [tbh]
\begin{center}
\includegraphics[width=8.6cm,height=9cm]{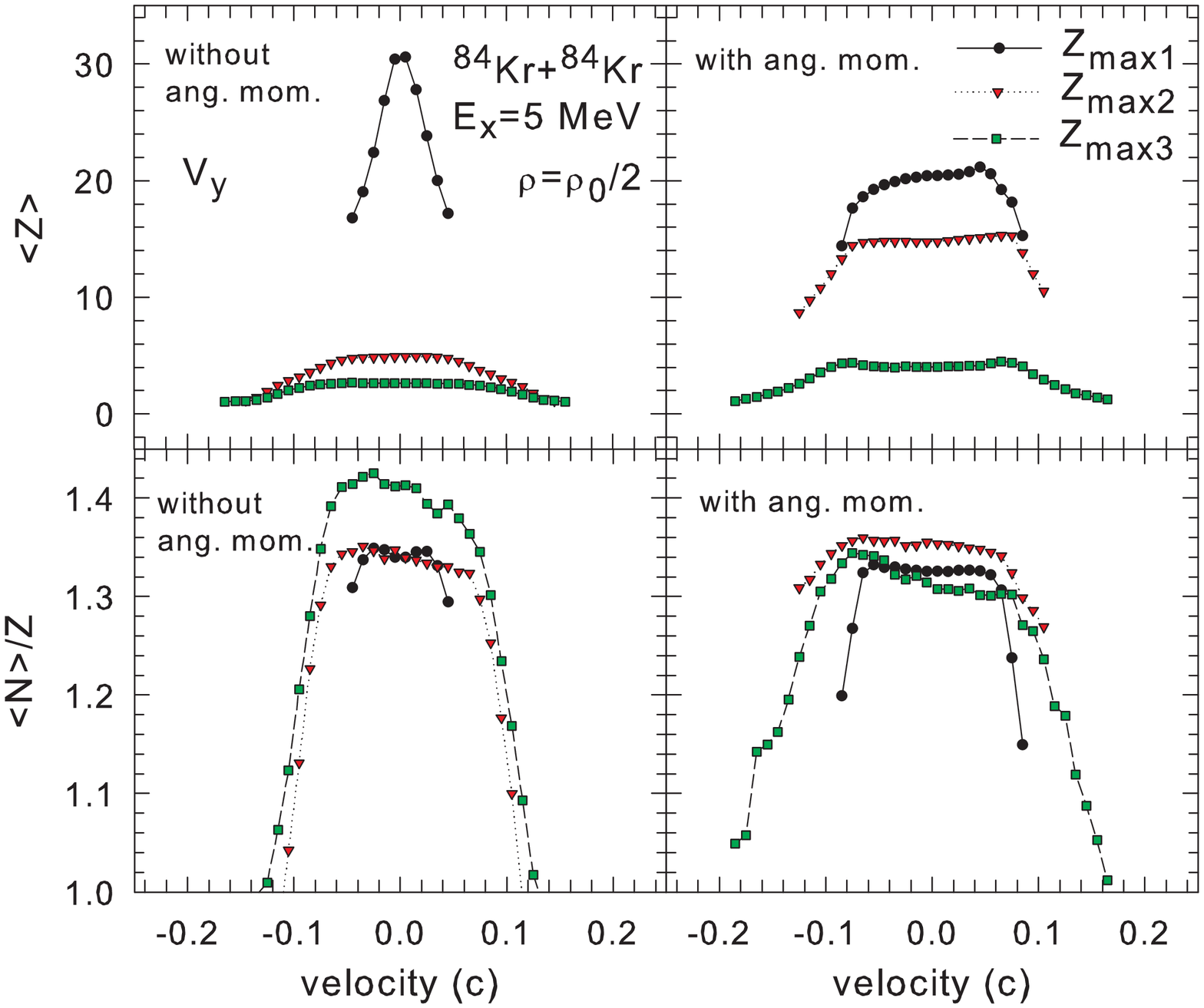}
\end{center}
\caption{\small{(Color online) Average charge $Z$ (top panels) and $<N>/Z$ 
(bottom panels) distributions of the first, second and third largest hot 
fragments ($Z_{max1},Z_{max2}$, and $Z_{max3}$) after multifragmentation 
of $^{84}$Kr projectile source at $E_x=5$ MeV/nucleon versus $V_y$ in the 
source frame. The freeze-out density is $\rho=\rho_0/2$. Left panels 
represent the results without angular momentum, while the right panels with 
angular momentum ($L=80\hbar$).}}
\end{figure}

\section{Effects of angular momentum and Coulomb interaction of projectile and 
target like sources}

As was mentioned, we analyze peripheral nucleus-nucleus collisions at 35 
MeV/nucleon with the corresponding relative velocity between the projectile 
and target about 80 mm/ns. 
At the initial dynamical stage of such a collision, the projectile nucleons 
interact with target nucleons and some energetic products of this 
interaction can leave the nuclei as pre-equilibrium particles. The 
kinetic energy of colliding nuclei can also be converted into the 
excitation energy of projectile and target residues. Therefore, the 
relative velocity between the residues decreases as well. These excited 
target and projectile-like sources decay afterwards. 

It is known that nuclear multifragmentation is a fast process within 
a characteristic time around 100 fm/c. Therefore, projectile and 
target-like sources will not be far from each other before disintegration. 
The idea is that at these short distances the long range Coulomb field of 
one of the sources influences the break up of the other one. 
In this case we are dealing with the multifragmentation in a double nuclear 
system, that is a new physical situation with respect to the standard 
multifragmentation of a single isolated source. 

According to our estimates from the energy conservation, their 
relative velocity should decrease to $\sim$50 mm/ns, at an 
excitation energy around 5 MeV/nucleon transferred to the residues. 
In this case, they will be separated
by $\approx$15 fm in a time of 100 fm/c. The decay of the 
two excited sources in such a double system is determined by the 
short-range nuclear forces. However, the presence of an external Coulomb 
field (for 
each source) can affect the composition of the produced 
fragments and their relative positions. In particular, an additional Coulomb 
barrier will prevent disintegration of the sources into many small pieces. 
It should be noted that during 
the evolution of a double system we must take into account its 
total center of mass conservation without a separate constraint in the 
freeze-out volumes of disintegrating sources. On the other hand, we 
include the angular momenta (rotation) of the separate sources, which 
can be transferred after the collision. It will also influence the positions 
and sizes of the fragments at the freeze-out 
\cite{BotvinaGross95,Botvina01,Botvina99}. 

In the following we demonstrate the results for multifragmentation of the 
projectile-like source (we call it the first source) by assuming the 
Coulomb field coming from the center of the target source (the second source). 
The first source is assumed to fly along $Y$-axis, and the second one is in 
the opposite direction (related to the center of mass of the double system). 
This separation axis may slightly deviate from the initial beam axis. The 
location of the second source is taken as $R_Y$= -10.6 fm and $R_Z$=10.6 fm 
with respect to the first source. The 
peripheral collision is assumed to take place in the $Y-Z$ plane, therefore, 
the coordinates in $Z$-axis is determined by the sizes of colliding 
nuclei, as well as by their possible repulsion after the collision. The X-axis 
is assumed to be an angular momentum axis. We suggest that this relative space 
configuration of the sources is quite general and suitable for investigating 
the Coulomb and angular momentum effects. 

The pre-equilibrium emission of few nucleons during the dynamical stage 
may decrease the excitation energy and relative velocity of the residues. 
However, this can be accounted in the statistical approach by changing 
the corresponding input and by using the ensemble of the sources 
(see, e.g., \cite{Ogul11}) with adequate parameters. On the other hand, 
as was shown by many theoretical and experimental works (see, e.g., 
\cite{bondorf95,Buyukcizmeci05,Milazzo00}), 
the relative yields of IMF do not depend much on the size of the sources 
in the multifragmentation regime (a scaling effect). Therefore, 
for our purposes it is sufficient to 
consider the sources of the same size and isospin content as the 
colliding nuclei ($^{84}$Kr).

\subsection{Charge Distributions}

The purpose of our analysis is to understand the new characteristics of 
fragment distributions, which are important for interpretation of 
many experiments on heavy-ion 
collisions at Fermi energies. We shall study the angular momentum and 
Coulomb field influences on the charge and isospin contents of produced 
fragments, and compare them with the standard calculations without these 
effects. It is expected that the correlations of the sizes and $<N>/Z$ of hot 
fragments with their velocities will be important. For this purpose, after 
the break-up of the sources we calculate the Coulomb propagation of 
produced hot fragments by taking into account the Coulomb interactions of 
particles in the double system and their velocities at the break-up time. 
In this paper, in order to clarify the modification of the multifragmentation 
picture caused by the new effects, we do not apply the secondary de-excitation 
of the hot fragments which, however, can lead to important consequences 
especially for isospin composition of final fragments. 

In Fig.~1, we show the total charge yields of hot fragments in case of 
with and without angular momentum conservation. Angular momentum value of 
$80 \hbar$ is selected as an upper 
estimate. It is seen that the charge distributions are very sensitive to the 
freeze-out densities. An angular momentum favors emission of large nearly 
symmetric fragments (like a nuclear fission) since the system in the 
freeze-out needs to have a large moment of inertia in order to minimize the 
rotational energy and to maximize the entropy. It is in a competition with 
the second source through the Coulomb interaction which prevents to emit an 
IMF with a large charge number. The latter can be seen clearly 
in comparison with the case of the fragmentation of a standard single 
isolated source. 

For more details, we show in Fig.~2 the yield distributions of the first, 
second and third largest fragments versus their charge numbers $Z_{max1}, 
Z_{max2}$, and $Z_{max3}$, respectively. These observables are used to 
obtain a complementary information about the fragmentation pattern. Top 
panels include the long-range Coulomb contribution from the second source 
only, and bottom panels include, in addition, 
the angular momentum effects. It is obvious that the distributions of 
$Z_{max1},Z_{max2}$, and $Z_{max3}$ are ordered according to their size. 
In the case of including 
angular momentum (bottom panels), average value of $Z_{max1}$ decreases, while 
$Z_{max2}$ and $Z_{max2}$ show an increasing trend.
A small freeze-out density ($\rho=\rho_0/6$) leads to the smoother and broader 
distributions (sometimes Gaussian-like ones) caused by the less 
restricted population of the larger coordinate phase space.

\subsection{$<N>/Z$ and Velocity Distributions}

The initial value of neutron-to-proton ratio ($N/Z$) of the source $^{84}$Kr 
is 1.33. In Fig.~3, we show that angular momentum leads to increasing N/Z 
values of IMFs in the case of strongly asymmetric decay. It is also caused 
by the increasing moment of inertia of the system that favors a bigger phase 
space of the reaction \cite{Botvina01}. It is a very instructive trend which 
can be responsible for many isospin observables. 

In Fig.~4, we present the yields per event as a function of the velocities 
$V_x$ and $V_y$ for the fragments having first, second and third biggest 
charge numbers. The $V_x$ velocity distributions in the direction of the 
angular 
momentum are shown in the top panel. The first and second fragments 
are nearly peaked around $0$, and it means that they are emitted mostly in 
$V_y-V_z$ plane. This is very different from the case of isotropic 
statistical emission taking place without angular momentum, and which is 
sometimes simplicitly assumed as the only possibility for the statistical 
break-up. 
In our case all these fragments fly predominantly in the plane of rotation, 
even though the smallest fragments can deviate from this plane. Another 
important effect can be seen in the bottom panel of Fig.~4. We remind that 
the velocity $V_y$ determines the 
separation axis (close to the beam axis), where the sources move in opposite 
directions. One can see that there is an order in the emission of fragments 
of different sizes, such that the largest fragments have the largest 
velocity $V_y$ , which is bigger than those of the second and third ones. 
The arrows refer to the average values of the velocity distributions. 
The maximum fragments fly predominantly in the forward direction as $V_y>0$, 
while the second and third ones fly to backward direction as $V_y<0$ 
(i.e., to the direction of the target source), and in this way they may 
simulate the so-called 'midrapidity emission'. This is the consequence of 
the fragment coordinate positions occupied predominantly in the 
freeze-out, and this is caused mainly by the Coulomb repulsion of the second 
source. This effect is consistent with the 
previous experimental observations \cite{colin}. Its dynamical interpretation 
may also be possible, however, the contribution of the Coulomb interaction 
was not separated and not investigated up to now with dynamical models. 

For detailed examination of the characteristics of produced particles, we 
show in Fig.~5, 
the relative yields (normalized per total number of generated events) 
of the hot primary fragments with $Z=$3, 6, 9, 12, 15, and 18 versus their 
velocities along the separation axis $V_y$. As in Fig.~4, these velocities 
are calculated with respect to the projectile source, so that $V_y<0$ means 
an emission towards the midrapidity. The figure demonstrates clearly the 
predominant 'backward' emission for the light IMF with changing to 
'uniform-like' emission for 
larger fragments. It is very important to investigate the isospin content 
of these fragments. Their $<N>/Z$ ratio versus $V_y$ is demonstrated in 
Fig.~6. One can see in this figure that including angular momentum can lead 
to increasing $<N>/Z$ ratio for small fragments. The external Coulomb field 
leads to the emission of neutron-rich fragments towards the midrapidity. 
This trend is clearly seen for Z=3, Z=6, and Z=9. The secondary decay 
of such fragments may preserve the enhanced neutron content and lead to 
emission of final neutron-rich nuclei in the 'neck' direction. However, 
this explanation may not be obvious for very small species (Z$\loo$3) because 
of their low contribution in partitions with large angular momentum, and 
because their preequilibrium emission is also possible in the reactions. 
Large species (Z$\goo$15) loose 
this sensitivity and they are emitted more uniformly. Including angular 
momentum increases also velocities of all fragments, and the distributions 
become broader. The drop of $<N>/Z$ at high velocities, actually, for 
fragments with low yields, is trivially explained by strong Coulomb 
acceleration if the mass is small. 

For completeness, we present how the average charge $<Z>$ (top panels) and 
$<N>/Z$ of the fragments with $Z_{max1},Z_{max2}$, and $Z_{max3}$ change 
with $V_y$ in Fig.~7, for the cases of with and without angular momentum 
effect. It is in agreement with our previous conclusions on modification 
of statistical picture. In particular, one can see a trend of increasing 
$<N>/Z$ towards the midrapidity for $Z_{max2}$ and $Z_{max3}$ 
(bottom panels). 

To verify our new-found trends we also performed the same calculations 
for heavier systems, e.g., $^{197}$Au+$^{197}$Au collisions. 
In all cases, we have got the same qualitative modifications of the standard 
multifragmentation picture.

\section{Relation between statistical and dynamical descriptions}

In this paper we investigate the kinematic characteristics, sizes and 
isospin properties of hot fragments. Subsequently, these fragments will be 
de-excited by emission of light particles, or by the secondary break-up, 
during their propagation. As was shown in many previous works the 
secondary process can be reliably described  within statistical models 
(see, e.g., Refs.~\cite{bondorf95,Ogul11} and references in). 
One of the outstanding problems under discussion is if dynamical 
models alone can describe the same evolutionary scenario 
leading to equilibration and multifragmentation as assumed by 
statistical models. In other words, is it possible to use only 
a dynamical description, instead of subdividing the whole reaction 
into dynamical and statistical stages? Some dynamical approaches try 
to reach this goal starting from 'first principles' like Fermionic 
molecular dynamics (FMD) \cite{FMD}, and antisymmetrized molecular 
dynamics (AMD) \cite{AMD}. Other approaches, like QMD \cite{QMD} and 
BNV \cite{BNV,colonna-piantelli} use semi-classical equations including 
two-body collisions 
and some elements of stochasticity. In all cases dynamical 
simulations are more complicated and time-consuming as compared 
with statistical models. This is why full calculations,
e.g. with FMD and AMD models, can only be done for relatively light 
systems. This prevents using these codes in many cases of 
nuclear fragmentation. A natural solution of this problem is to 
develop hybrid approaches which combine dynamical models 
for describing the nonequilibrium early stages of the 
reaction with statistical models for describing the fragmentation 
of equilibrated sources. In this respect, the statistical and 
dynamical approaches are complementary and suitable for many practical 
calculations, which are required, e.g., in medicine, space research, 
and other fields. 

One can try dynamical model to describe the fragment (i.e., nucleon 
cluster) formation at the time of the nuclear freeze-out 
($\sim$100 fm/c) too, as we are doing for hot fragments with statistical 
models. Actually, it is popular to explain in a dynamical way 
the formation of neck-like fragments in collisions of heavy ions  
around the Fermi energy \cite{BNV}. 
However, the essential problem in the dynamical approach is the connection 
to the relatively slow secondary de-excitation stage of the fragment 
($\goo 10^2 - 10^3$ fm/c).
This last stage is very important 
for isotope composition of final cold fragments, which may give access 
to the symmetry energy of nuclei and nuclear matter. As shown in some 
dynamical calculations \cite{BNV}, the primary nucleon clusters may have 
a low density and unusual form. So it is difficult to establish 
an excitation energy of such clusters \cite{Liu04}. Moreover, a bigger 
problem is the evaluation of other properties of these clusters, such 
as their masses, level densities, and symmetry energies. The last ones 
are crucially important for the subsequent de-excitations leading to cold 
nuclei. To our knowledge, nobody has ever described realistically 
these properties within transport dynamical models. Therefore, the 
de-excitation results 
for isotope composition of final nuclei become not very reliable, and, 
usually, the predictions of dynamical models are limited by hot fragments. 
On the other hand, one can easily resolve this problem within the 
statistical approach. As demonstrated in 
Refs.~\cite{LeFevre05,Ogul11,Buyukcizmeci05} one can connect the 
freeze-out properties of hot fragments with their secondary de-excitation 
and the yield of final isotopes. In this respect, the application of 
appropriate statistical models to the reactions, which were considered 
previously only as dynamical processes, opens real chances for involving 
new data in theoretical analysis. One should bear in mind that the 
statistical and dynamical approaches are derived from different physical 
principles. The time-dependent dynamical approaches are 
based on Hamiltonian dynamics (the principle of minimal 
action), whereas the statistical models employ the principle 
of uniform population of the phase space. Actually, 
these two principles represent complementary methods for describing 
the physical reality. Therefore, a decision of using statistical or 
dynamical approaches for the description of nuclear multifragmentation 
should be made after careful examination of the degrees of equilibration 
expected in particular cases, and it can only be justified by the 
comparison with experiment.

In the case of equilibrated sources, the predictions of statistical 
models are usually in better agreement with experimental data. 
This is well known in multifragmentation of relativistic 
projectiles \cite{botvina95,xi97,EOS}, especially when the chemical 
equilibrium is established in such reactions, and this equilibrium can 
be seen in isotopic yields \cite{LeFevre05,Ogul11}. 
This can also be seen by describing the isospin observables in 
nucleus-nucleus collisions at lower (Fermi) energies: For example, one can 
compare dynamical \cite{Liu04} and statistical 
\cite{Ogul09,Buyukcizmeci12,DasGupta} analyses of the 
MSU experimental data. As shown in the present work, the effect 
of increasing the neutron number in IMF emitted towards the 
midrapidity region may also be explained within the statistical picture 
modified by including the external Coulomb field and angular momentum. 
In order to distinguish dynamical and statistical mechanisms one should 
involve specific isotopic experimental characteristics. For example, 
there are experimental data demonstrating the trend of 
increasing neutron richness of IMF in 
collisions of nuclei with increasing centrality, i.e. with increasing 
excitation energy, both in the 'neck region' \cite{Xu02} and in the 
single equilibrated central source \cite{Milazzo00}. Both trends can 
easily be explained within the statistical framework 
\cite{Botvina01,Buyukcizmeci05,Milazzo00} 
because of the enhanced disintegration of nuclei at high excitations into 
more neutron-rich IMF, while a dynamical calculation predicts more 
neutron-rich IMF in very peripheral collisions since the 
neutron-rich periphery of nuclei influences the dynamics of IMF 
formation \cite{BNV}. We believe that the presented generalization of 
the statistical approach is very useful for the analysis of coming novel 
experiments 
aimed predominantly at measuring isotopes at low and intermediate energy 
collisions \cite{fazia13, staggering}. As it was extensively discussed 
in Ref.~\cite{Gross97}, 
some 'dynamical' behaviour of many-body systems constraint by the 
conservation laws and influenced by the long-range forces may be simulated 
within the microcanonical statistical ensemble.

\section{Conclusions}

In conclusion, within the statistical approach we have investigated isotopic 
characteristics of hot fragments after the multifragmention of the Kr-like 
projectiles in peripheral $^{84}$Kr+$^{84}$Kr collisions around the Fermi 
energy. It is important and new that we have taken into account Coulomb and 
angular momentum effects originated after the collision dynamics. 
We used the microcanonical Markov chain approach within the statistical 
multifragmentation model. 
It is shown that conservation of angular momentum and complicated Coulomb 
interactions caused by the proximity of target and projectile-like sources in 
the freeze-out stage produce significant changes in the multifragmentation 
picture. There appear new fragment formation trends, such as an asymmetry of 
IMF emission (predominantly towards the midrapidity), increasing the neutron 
content of these IMF, a correlation (ordering) of sizes and velocity of 
fragments, and in-plane emission of large fragments. 
This is instructive since in previous years it was assumed that such effects 
could be explained within dynamical models only. These features may also be 
preserved after the secondary excitation of hot fragments for the cold 
fragments, similar to the previously analyzed reactions leading to the 
production and decay of the single isolated sources. For the future, we plan 
to apply this new approach 
to analyze the experimental data at intermediate peripheral collisions such 
as the FAZIA data measured in $^{84}$Kr+$^{124,124}$Sn reactions at 
35 Mev/nucleon \cite{fazia13}. Particular isotopic effects, such as the 
odd-even staggering of the yield of final fragments studied by FAZIA 
collaboration  \cite{staggering}, can also be analysed within similar 
statistical approaches. Some preliminary encouraging results obtained with 
the help of the ensemble of residual 
sources were already reported \cite{nufra2013}. 
This kind of investigations will show a new connection between dynamical 
and statistical phenomena in nuclear reactions. As expected, it may also 
provide us with inputs to understand the 
nuclear equation of state and nuclear composition, which are important to 
determine the properties of nuclear and stellar matter at extreme conditions 
and their connections to the thermodynamics of stellar matter in astrophysical 
events \cite{nihal13}. We believe that our theoretical results may be 
enlightening for further analysis of the experiments. \\

\begin{acknowledgments}
This work was supported by TUBITAK (Turkey) with project number 113F058.
We thank G. Casini for stimulating discussions and help in the preparation 
of the manuscript. A.S.B. acknowledges also a support by HIC for 
FAIR (LOEWE program). 
\end{acknowledgments}

\end{document}